\documentclass[10pt,twocolumn,letterpaper,superscriptaddress,prl]{revtex4}

\usepackage{graphicx,amsmath,color,amssymb}
\usepackage[latin1]{inputenc}
\usepackage[sort&compress]{natbib}
\usepackage{epstopdf}
\usepackage{booktabs}
\usepackage[greek,english]{babel}
\bibpunct{}{}{,}{s}{}{,}

\usepackage[dvips]{epsfig}
\usepackage{subfigure,psfrag}
\usepackage{lscape}
\usepackage{varioref}
\usepackage{float}
\usepackage{tabularx}
\usepackage[version-1-compatibility]{siunitx}
\definecolor{darkblue}{rgb}{0,0,.4}
\definecolor{darkred}{rgb}{0.6,0,0}

\newcommand{\abs}[1]{\left|#1\right|}

\newcommand{\ket}[1]{\left| #1 \right>}
\newcommand{\bra}[1]{\left< #1 \right|}
\newcommand{\bracket}[3]{\left< #1 \abs{#2} #3 \right>}

\newcommand\figref[1]{Fig.~\ref{#1}}
\renewcommand\eqref[1]{Eq.~(\ref{#1})}

\newcommand{\rr}{\mathbf{r}}

\newcommand{\ovI}[2]{\left< #1 | #2 \right>}

\usepackage[bookmarksopen=true, 
citecolor=blue,
colorlinks=true, 
linkcolor=blue]{hyperref} 

\begin{document}

\title{Probing electric and magnetic vacuum fluctuations with quantum dots}

\author{P.~Tighineanu}
\email{petru.tighineanu@nbi.ku.dk}
\affiliation{Niels Bohr Institute,\ University of Copenhagen,\ Blegdamsvej 17,\ DK-2100 Copenhagen,\ Denmark}

\author{M.~L.~Andersen}
\affiliation{Niels Bohr Institute,\ University of Copenhagen,\ Blegdamsvej 17,\ DK-2100 Copenhagen,\ Denmark}

\author{A.~S.~S\o rensen}
\affiliation{Niels Bohr Institute,\ University of Copenhagen,\ Blegdamsvej 17,\ DK-2100 Copenhagen,\ Denmark}

\author{S.~Stobbe}
\affiliation{Niels Bohr Institute,\ University of Copenhagen,\ Blegdamsvej 17,\ DK-2100 Copenhagen,\ Denmark}
\author{P. Lodahl}
\email{lodahl@nbi.ku.dk}
\homepage{http://www.quantum-photonics.dk/}
\affiliation{Niels Bohr Institute,\ University of Copenhagen,\ Blegdamsvej 17,\ DK-2100 Copenhagen,\ Denmark}

\date{\today}

\small

\begin{abstract}
The electromagnetic-vacuum-field fluctuations are intimately linked to the process of spontaneous emission of light. Atomic emitters cannot probe electric- and magnetic-field fluctuations simultaneously because electric and magnetic transitions correspond to different selection rules. In this paper we show that semiconductor quantum dots are fundamentally different and are capable of mediating electric-dipole, magnetic-dipole, and electric-quadrupole transitions on a single electronic resonance. As a consequence, quantum dots can probe electric and magnetic fields simultaneously and can thus be applied for sensing the electromagnetic environment of complex photonic nanostructures. Our study opens the prospect of interfacing quantum dots with optical metamaterials for tailoring the electric and magnetic light-matter interaction at the single-emitter level.
\end{abstract}


\maketitle

Spontaneous emission is a fundamental physical process, which plays an essential role in nature as the main source of optical radiation, and in applications as the principal source of artificial illumination. Quantum mechanically, spontaneous emission is an effect of the fluctuating electromagnetic vacuum field perturbing the emitter. At optical frequencies, emitters sense mainly the electric field while higher-order multipole field components can be neglected. This is because the variation of the electromagnetic field is negligible over the spatial extent of most quantum emitters, which has rendered the dipole approximation a highly successful approximation in quantum electrodynamics. Nevertheless, magnetic-dipole (MD) and electric-quadrupole (EQ) transitions are well known in atomic physics and can be accessed with light despite being much weaker,\cite{noecker88,rukhlenko09,taminiau12} since they have different selection rules than electric-dipole (ED) transitions.\cite{zurita01,zurita02} Semiconductor quantum dots (QDs) are however fundamentally different. Unlike atoms, the dipole approximation may not apply to QDs even on dipole-allowed transitions.\cite{andersen10} The asymmetry of the QD wavefunctions originating from a lack of mirror-reflection symmetry (parity symmetry) of the QD confinement potential breaks the usual selection rules applicable in atomic physics leading to both ED and MD contributions on the same transition. For atoms a related but very weak asymmetry is induced by the electroweak interaction and has been used to probe the standard model of particle physics.\cite{wood97} In contrast, the parity violation is very strong for QDs due to their asymmetric structure and, therefore, they may be exploited as a sensitive probe of the parity conservation and the nature of the multipolar quantum-vacuum fluctuations in complex photonic nanostructures.

\begin{figure}[t!]
	\includegraphics[width=0.23\textwidth]{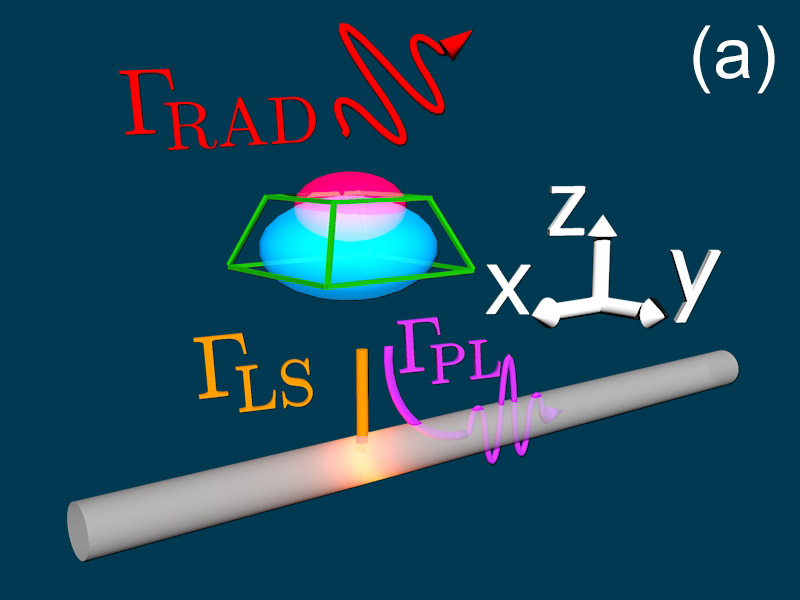} \includegraphics[width=0.23\textwidth]{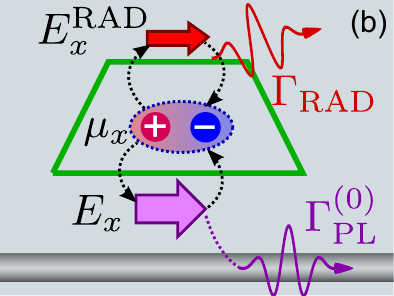}\\[0.05cm]
	\includegraphics[width=0.23\textwidth]{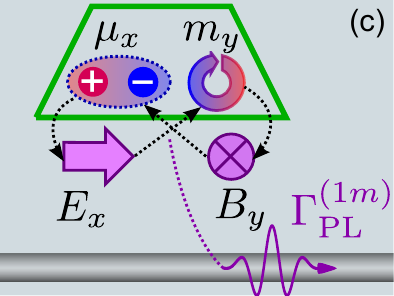} \includegraphics[width=0.23\textwidth]{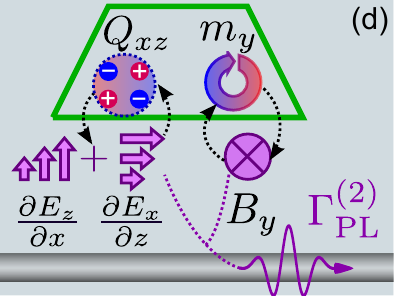}
	\caption{ \label{fig:fig1} (Color online) Decay dynamics of mesoscopic QDs beyond the dipole approximation. (a) Schematic of the QD decay channels in the proximity of a metal interface. The electron (blue) and hole (red) wavefunctions illustrate the built-in asymmetry. (b) Light-matter interaction processes governing $\Gamma^{(0)}$, where the ED interacts with the radiation modes of the electric vacuum $E_x^\mathrm{RAD}$  and the guided surface plasmon modes $E_x$. (c) Processes governing the ED-MD interference. The light emitted by the ED $\mu_x$ interacts with the MD $m_y$ and creates a magnetic field. The physical picture of $\Gamma^{(1Q)}$ is conceptually analogous. (d) Processes governing $\Gamma^{(2)}$ with pure MD and EQ contributions. The EQ $Q_{xz}$ couples to the gradient of the electric vacuum.}
\end{figure}

In the present work, we show that the commonly used self-assembled In(Ga)As QDs have ED, MD, and EQ interactions of comparable magnitude on a single electronic resonance, which is a new and fundamental effect in the fields of nano optics and quantum optics. A current hot topic in nanophotonics exploits the role of non-locality of the dielectric response in plasmonics.\cite{ciraci12,toscano13} Here we study a different non-local phenomenon by accounting for the spatial extent and symmetry of QDs in nanostructures of importance for photon emission. The effect is particularly pronounced, if both the QD and the nanophotonic environment violate parity symmetry. The developed formalism is remarkably simple, as we obtain a single light-matter interaction channel for the multipolar part, which, combined with the usual electric-dipole contribution, describes completely the QD-field interaction. The effect is important for many nanophotonic configurations but for concreteness we consider the QD spontaneous emission for two experimentally realistic nanophotonic structures: a semiconductor-metal plane interface and a plasmonic nanowire (see Fig.~\ref{fig:fig1}(a)). Our novel theoretical model employs simple parity arguments to identify the relevant electric and magnetic contributions in a multipolar field expansion. A similar underlying concept is used in the description of the absorption of X-rays by molecules\cite{bernadotte12} for the case of plane-wave illumination but is incompatible with the complex field profiles encountered in photonic nanostructures. Our study demonstrates that single QDs can be employed for locally probing complex photonic nanostructures that tailor both the electric and magnetic field.\cite{soukoulis10,pendry06} Sensitivity to the magnetic field has been a long-sought goal in nanophotonics, and has been achieved so far only by scanning near-field spectroscopy\cite{burresi09} where the disturbance of the electromagnetic field profile by the applied near-field probe can be an issue. The nanometer-size of single QDs enables non-invasive probing that operates at the single-electron single-photon level. Furthermore, the multipolar coupling of QDs can potentially be exploited for enhancing the light-matter interaction with immediate applications to quantum light sources for quantum-information processing.\cite{lodahl13}

The QD spontaneous-emission rate $\Gamma$ quantifies the light-matter interaction strength and is often used as the experimental signature of the interaction.\cite{wang11,tighineanu13} According to Fermi's Golden Rule,\cite{tannoudji05}  
\begin{equation}
\Gamma=\frac{\pi}{\epsilon_0\hbar \omega}\sum_l\abs{T_l}^2\delta(\omega-\omega_l),
\label{eq:Gamma_def}
\end{equation}
where the generalized Coulomb gauge\cite{vats02} is applied and $e$ and $m_0$ are the elementary charge and electron mass, $T_{l}=e/m_0\bracket{\Psi_g}{\mathbf{f}_l^*(\mathbf{r})\cdot\hat{\mathbf{p}}}{\Psi_e}$ is the transition matrix element between the ground $\ket{\Psi_g}$ and excited $\ket{\Psi_e}$ electronic states of the QD, $\hat{\mathbf{p}}=-i\hbar\nabla$ is the momentum operator and $\mathbf{f}_l$ is the normal vector-potential mode.\cite{novotny12}
The sum over $l$ indicates that the emitter can decay into an infinite number of spatial modes at the resonance frequency $\omega$. At this point, the standard textbook approach is to invoke the dipole approximation $\mathbf{f}_l(\rr)\approx\mathbf{f}_l(\rr_0)$ by assuming that the field varies slowly over the spatial extent of the QD $(L_\mathrm{QD})$, i.e., $kL_\mathrm{QD}\ll 1$, where $k$ is the wavevector of light and $\mathbf{r_0}$ the position of the QD. The decay of such (small) emitters is triggered solely by the electric vacuum field, and the MD and EQ components do not play any role.

\begin{figure*}[t!]
	\includegraphics[height=0.16\textheight]{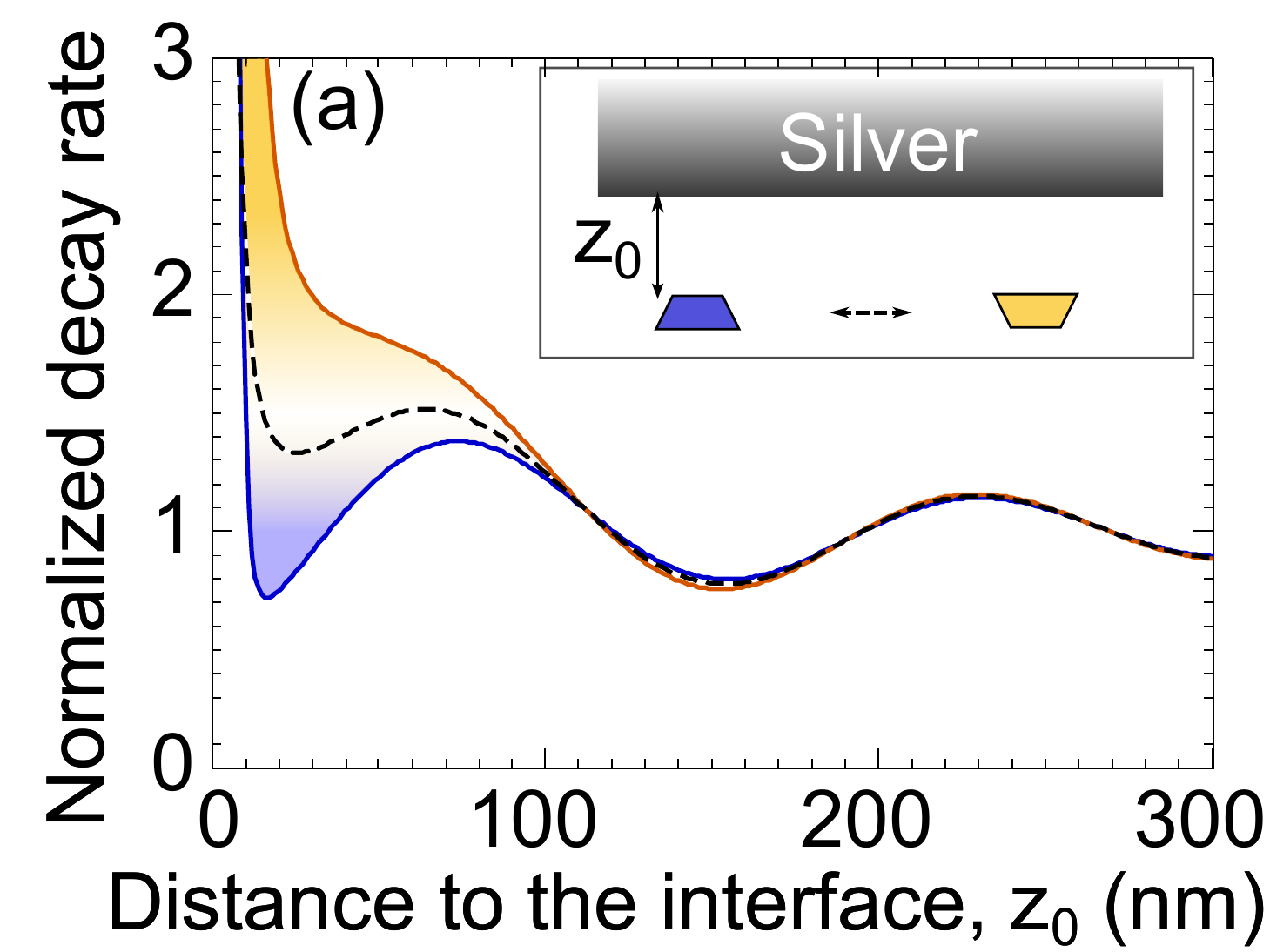} \includegraphics[height=0.16\textheight]{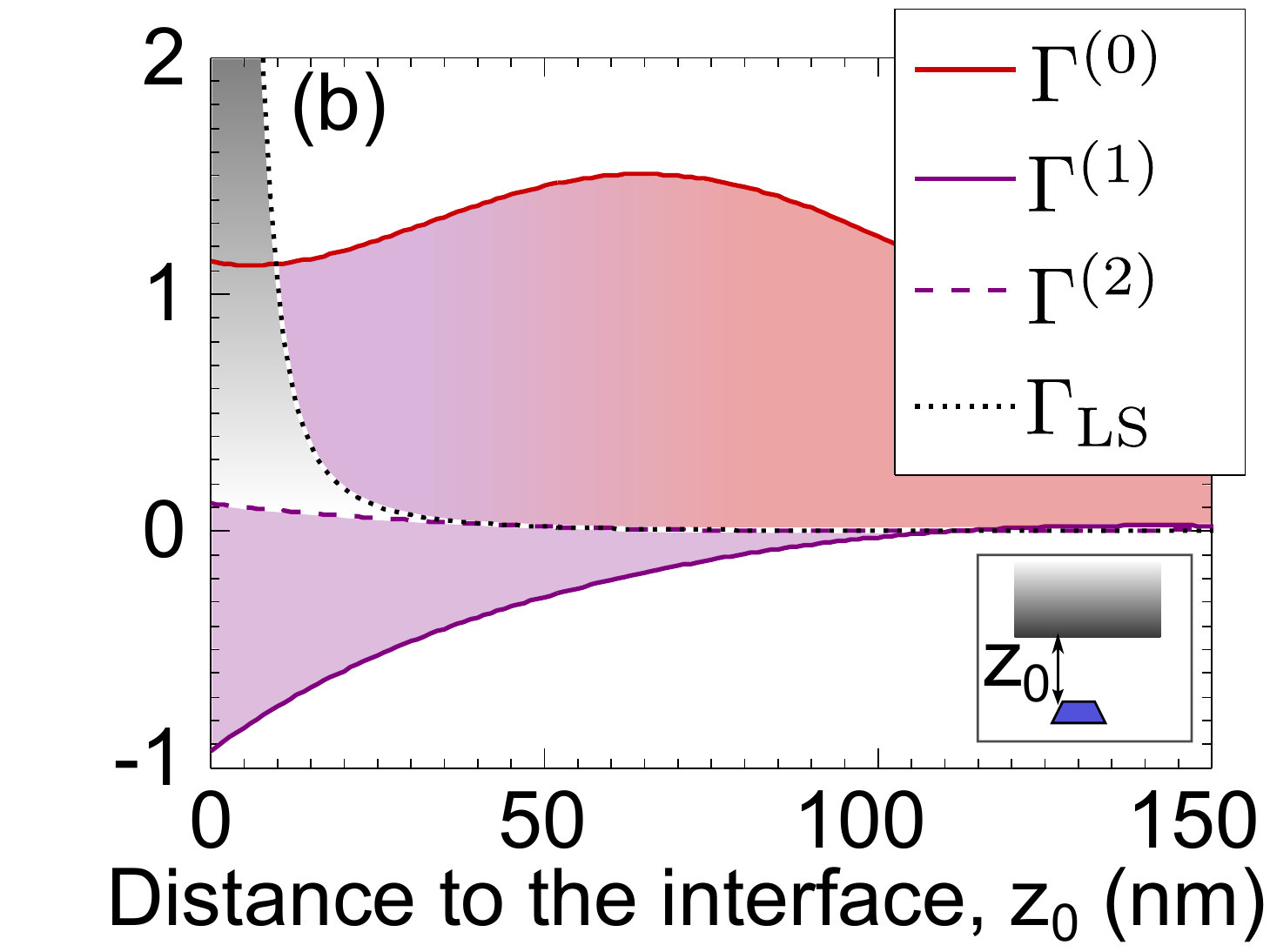} \hspace{0.3cm} \includegraphics[height=0.16\textheight]{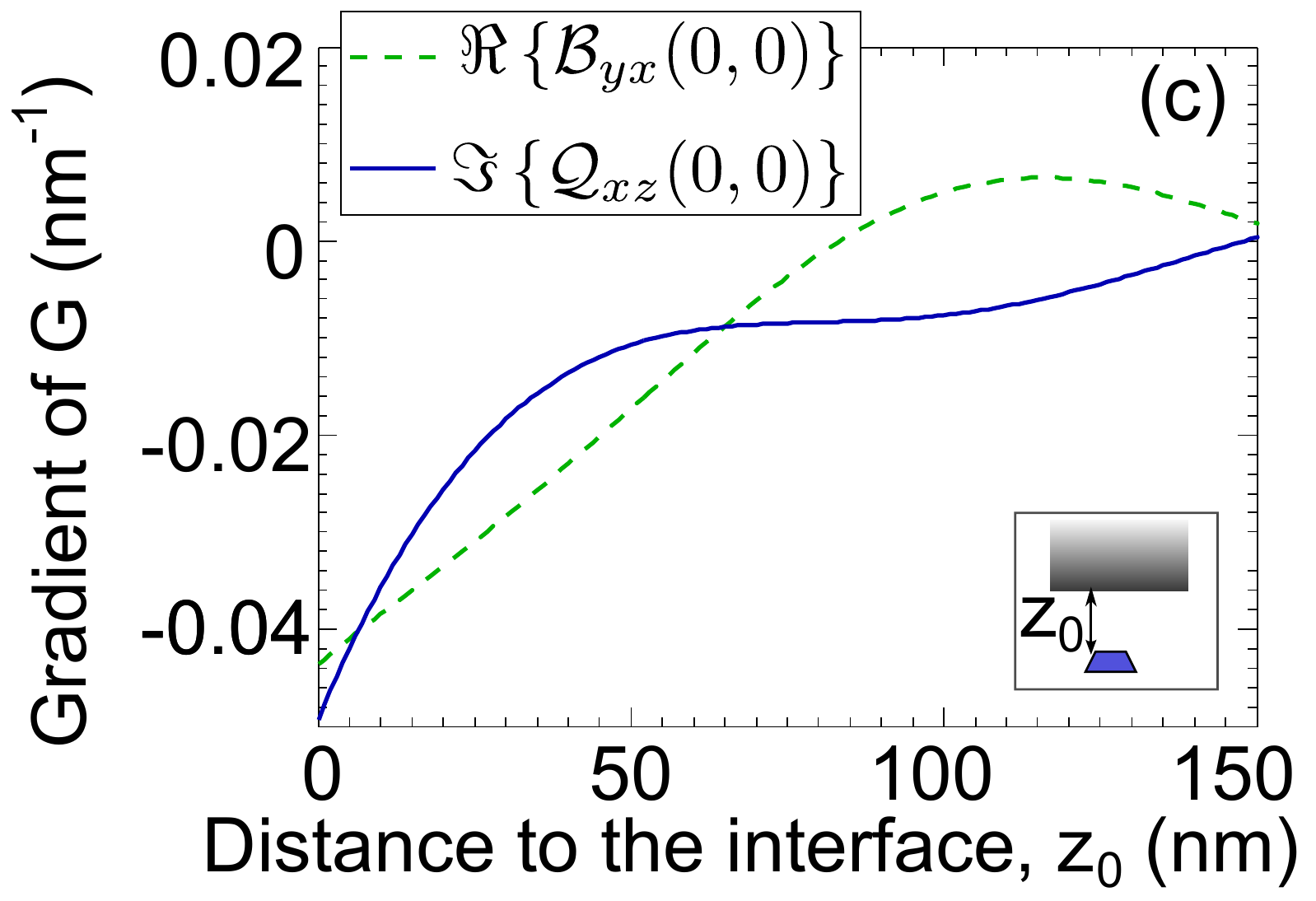}
	\caption{ \label{fig:fig2} (Color online) Decay dynamics of In(Ga)As QDs near a silver interface. All the rates are normalized to the decay rate in homogeneous GaAs. (a) Decay rate for the direct (inverted) QD orientation marked by blue (orange) solid lines. The black dashed line corresponds to the dipole theory. (b) Decomposition of the decay rates according to the expansion order. The near-field ohmic losses are indicated by the dotted black line. (c) The ED-MD and ED-EQ Green's tensor probed by mesoscopic QDs. The quantity is normalized to $\Im\{ G_{xx}(0,0) \}$ in homogeneous GaAs.}
\end{figure*}

Andersen et al\cite{andersen10} showed experimentally that QDs do not always behave as point dipoles in spontaneous-emission experiments. Therefore, we account for the field varying over the QD and perform a Taylor expansion in the field modes $f_{l,i}(\rr)=f_{l,i}(0)+x_j\partial_jf_{l,i}(0)+x_jx_k\partial_k\partial_jf_{l,i}(0)/2+...$, where we use the summation convention over repeated indices and choose $\rr_0=0$ without loss of generality. We note that the expansion can be performed as long as $kL_\mathrm{QD}<1$, otherwise \eqref{eq:Gamma_def} must be evaluated directly,\cite{ahn03} complicating the problem and offering limited physical insight.~\cite{stobbe12,kristensen13} The expansion is inserted into the transition matrix element and the corresponding orders are collected accordingly
\begin{equation}
\begin{split}
T_{l} &= T_l^{(0)}+T_l^{(1)}+T_l^{(2)}+...\\
&= \mu_if_{l,i}^*(0) + \Lambda_{ji}\partial_jf_{l,i}^*(0) + \Omega_{kji}\partial_j\partial_kf_{l,i}^*(0)+...,
\end{split}
\label{eq:T_multipolar}
\end{equation}
where $\mu_i=e/m_0\bracket{\Psi_g}{\hat{p}_i}{\Psi_e}$, $\Lambda_{ij}=e/m_0\bracket{\Psi_g}{x_i\hat{p}_j}{\Psi_e}$, and $\Omega_{ijk}=e/2m_0\bracket{\Psi_g}{x_ix_j\hat{p}_k}{\Psi_e}$ are the dipole moment, first-order, and second-order mesoscopic moments of the QD, respectively. $\mathbf{\Lambda}$ ($\mathbf{\Omega}$) is a two-dimensional (three-dimensional) tensor and contains magnetic-dipole (magnetic-quadrupole) and electric-quadrupole (electric-octupole) contributions.\cite{bernadotte12} In the Supplementary Information we analyze the symmetry properties of the wavefunctions and show that $\boldsymbol \mu$ and $\boldsymbol \Lambda$ have the form
\begin{equation}
\boldsymbol\mu =
\begin{pmatrix}
\mu \\ 0 \\ 0
\end{pmatrix}
,\hspace{.5cm}
\mathbf{\Lambda}=
\begin{pmatrix}
0 & 0 & \Lambda\\
0 & 0 & 0\\
0 & 0 & 0
\end{pmatrix},
\label{eq:mesMoments}
\end{equation}
where $\mu\equiv \mu_x$ and $\Lambda\equiv\Lambda_{xz}$. Furthermore, we show that the contribution of $\mathbf{\Omega}$ can be neglected. Remarkably, one single parameter, $\Lambda$, describes the light-matter interaction beyond the dipole approximation and contains MD and EQ moments as will be shown later. The ratio $\abs{\Lambda/\mu}$ quantifies the mesoscopic character of the QD and was measured to be about \SI{10}{\nano\meter} for standard self-assembled In(Ga)As QDs.\cite{andersen10} This value will be used throughout the paper. The interaction with light can be either suppressed or enhanced by the mesoscopic moment $\Lambda$ depending on the properties of the environment of the QD. This effect is illustrated in Fig.~\ref{fig:fig2}(a), where we show the emission rate of a QD in the proximity of a silver interface (at an emission wavelength of $\SI{1000}{\nano\meter}$  and with the refractive indices of GaAs $n_\mathrm{GaAs}=3.42$ and of silver $n_\mathrm{Ag}=0.2+7i$). Note that, unlike QDs, atomic wavefunctions possess parity symmetry so that $\boldsymbol \mu$ and $\boldsymbol \Lambda$ never contribute simultaneously.

We expand the decay rate as $\Gamma \approx \Gamma^{(0)} + \Gamma^{(1)} + \Gamma^{(2)} \propto (T_l^{(0)}+T_l^{(1)}+T_l^{(2)})^*(T_l^{(0)}+T_l^{(1)}+T_l^{(2)})$. In the proximity of metals, the QD can decay into propagating photons with the rate $\Gamma_\mathrm{RAD}$, propagating surface plasmons $(\Gamma_\mathrm{PL})$, or lossy modes in the metal $(\Gamma_\mathrm{LS})$, see Fig.~\ref{fig:fig1}(a). The former coupling to radiative modes is essentially not affected by multipolar effects since the responsible electromagnetic field varies weakly in space, i.e., $\Gamma_\mathrm{RAD} \approx \Gamma_\mathrm{RAD}^{(0)}$. In contrast, the plasmon field varies strongly and therefore multipolar effects influence the excitation rate of plasmons. The coupling to lossy modes is normally negligible for distances larger than $\sim \SI{20}{\nano\meter}$ from the metal and we therefore do not discuss them further. We combine Eqs.~(\ref{eq:Gamma_def}--\ref{eq:mesMoments}) and obtain the three light-matter interaction channels for mesoscopic QDs
\begin{equation}
\begin{split}
\Gamma^{(0)} &= A\abs{\mu}^2 \Im\left\{ G_{xx}(0,0) \right\}= \Gamma_\mathrm{RAD}+\Gamma_\mathrm{PL}^{(0)}\\
\Gamma^{(1)} &= 2A\Re\left(\Lambda\mu^*\right) \left. \partial_x \Im\left\{ G_{zx}(\rr,0) \right\}\right|_{\rr=0}\approx\Gamma_\mathrm{PL}^{(1)}\\
\Gamma^{(2)} &= A\abs{\Lambda}^2 \left. \partial_x \partial_x' \Im\left\{ G_{zz}(\rr,\rr') \right\} \right|_{\rr=\rr'=0}\approx \Gamma_\mathrm{PL}^{(2)},
\end{split}
\label{eq:Gamma_orders}
\end{equation}
where $A=2e^2/\epsilon_0\hbar m_0^2c_0^2$, $c_0$ is the vacuum speed of light, and $\Re$ and $\Im$ denote the real- and imaginary-part operators, respectively. We have defined the Green's tensor $G_{ij}(\rr,\rr';\omega)=(\pi c_0^2/2\omega^3)\sum_le_{l,i}^*(\rr)e_{l,j}(\rr')\delta(\omega-\omega_l)$\cite{paulus00} and $\mathbf{e}_{l}(\rr)=i\omega \mathbf{f}_l(\rr)$ is the normal electric-field mode. Each order has a clear physical meaning as explained below, where we exemplify a semiconductor-silver interface as sketched in the inset of Fig.~\ref{fig:fig2}(a).

The zeroth-order rate, $\Gamma^{(0)}$, is the well-known ED contribution, and is given as a product of a field term, $\Im\left\{ G_{xx} \right\}$, which is proportional to the (electric) local density of optical states, and a QD term, $\abs{\mu}^2$, which is proportional to the (electric) oscillator strength.\cite{novotny12} Here, a microscopic polarization in the $x$-direction couples to the $x$-polarized electric field, which probes the environment and interferes back with itself. The resulting field excitation propagates away from the QD in the form of free photons or surface plasmons, see Fig.~\ref{fig:fig1}(b). In the proximity of an interface, $\Gamma^{(0)}$ has the well-known Drexhage dependence,\cite{drexhage70} see Fig.~\ref{fig:fig2}(a,b), where the red-violet color gradient indicates that the coupling to the plasmonic field becomes dominant at distances smaller than $\sim\SI{50}{\nano\meter}$.

The higher-order corrections to $\Gamma$ depend on the mesoscopic moment $\Lambda$, which is responsible for the non-local interaction with light. It is clear from \eqref{eq:Gamma_orders} that $\Gamma^{(1)}$ is a first-order process and is negligible, if the figure of merit $\mathcal{G}^{(1)}\equiv |\Gamma^{(1)}|/\Gamma^{(0)}\approx k\times 2\abs{\Lambda/\mu}$ is much smaller than unity. For In(Ga)As QDs, $\mathcal{G}^{(1)}\simeq 0.44$ shows that the light-matter interaction beyond the dipole approximation can be strong. The magnitude of such effects is determined by the field gradients of the particular photonic nanostructure and we compute them in the next paragraph. $\Gamma^{(2)}$ is a second-order process and contains pure MD and EQ contributions as sketched in Fig.~\ref{fig:fig1}(d). For QDs, the important quantity is $\mathcal{G}^{(2)}\equiv |\Gamma^{(2)}|/\Gamma^{(0)}\approx k^2\abs{\Lambda/\mu}^2 \simeq 0.05$, which is negligible.  Note that the dipole approximation is more robust for atoms and other high-symmetry emitters, since the first non-vanishing contribution is $\Gamma^{(2)}$, which has a weight of $(kL_\mathrm{QD})^2$ with respect to $\Gamma^{(0)}$.  

In the following we discuss the first-order contribution, $\Gamma^{(1)}$, in quantitative terms. The mesoscopic moment contains MD and EQ contributions, as can be seen from
\begin{equation}
\Lambda_{xz}\partial_xe_{l,z}(0) =  i\omega m_yb_{l,y}(0) + Q_{xz}\left[\partial_xe_{l,z}(0)+\partial_ze_{l,x}(0)\right],
\label{eq:multipolar}
\end{equation}
where $\mathbf{b}=-i\omega^{-1}\nabla\times\mathbf{e}$ is the magnetic-field mode, $m_y\equiv m = (e/2m_0)\bracket{\Psi_g}{x\hat{p}_z-z\hat{p}_x}{\Psi_e}$ the MD, and $Q_{xz} \equiv Q = (e/2m_0)\bracket{\Psi_g}{x\hat{p}_z+\hat{p}_xz}{\Psi_e}$ the EQ of the QD. The two moments are equal, i.e., $m=Q=\Lambda/2$, but they couple to different field components and, therefore, their contribution can be tailored independently. As a consequence, $\Gamma^{(1)}$ intertwines the ED, MD and EQ characters of the QD with the following physical interpretation. The ED couples to the $x$-polarized electric field, which probes the environment and interferes back with the MD and EQ components, see Fig.~\ref{fig:fig1}(c). The resulting field excitation propagates away in the form of surface plasmons. Note that $\Gamma^{(1)}\neq 0$ only if both the QD wavefunctions and the electromagnetic environment violate parity symmetry. This is because a parity-symmetric electronic potential cannot be both $\boldsymbol \mu$- and $\boldsymbol \Lambda$-allowed, and a parity-symmetric environment contains either even or odd electromagnetic modes. The ED is an even operator and would couple only to the even modes, while $\Lambda$ corresponds to an odd operator and would couple to the odd modes inducing no mutual interference between $\boldsymbol \mu$ and $\boldsymbol \Lambda$ and a vanishing $\Gamma^{(1)}$. The first-order contribution can both enhance and suppress the light-matter interaction depending on whether the light emitted by the ED interferes constructively or destructively with the mesoscopic moment $\Lambda$. This can be seen in Fig.~\ref{fig:fig2}(a), where by flipping the QD orientation $\Lambda$ changes sign but the field gradient does not and, hence, $\Gamma^{(1)}$ changes from suppressing to enhancing the decay rate. The multipolar contribution to $\Gamma^{(1)}$ is
\begin{equation}
\begin{split}
\Gamma^{(1)} &= \Gamma^{(1m)} + \Gamma^{(1Q)}\\
             &= A\omega m_y\mu^*\Re\left\{\mathcal{B}_{yx}(0,0)\right\} + AQ_{xz}\mu^*\Im\{ \mathcal{Q}_{xz}(0,0) \},
\end{split}
\label{eq:G1MQ}
\end{equation}
where we define the ED-MD Green's tensor $\mathcal{B}_{yx}(0,0) = -i\omega^{-1}\left[\partial_xG_{zx}(\rr,0)-\partial_zG_{xx}(\rr,0)\right]_{\rr=0}$, the ED-EQ Green's tensor $\mathcal{Q}_{xz}(0,0)=\left[ \partial_xG_{zx}(\rr,0)+\partial_zG_{xx}(\rr,0) \right]_{\rr=0}$, and assume $\Lambda\mu^*$ to be real (this holds in the effective-mass approximation of the QD wavefunctions and will be discussed elsewhere\cite{tighineanu14}). \eqref{eq:G1MQ} shows that QDs access the magnetic and electric-quadrupole vacuum fields, similar to the way dipoles probe the electric component of the vacuum. This is demonstrated in Fig.~\ref{fig:fig2}(c), where the contribution of $\Re\left\{\mathcal{B}_{yx}(0,0)\right\}$ and $\Im\left\{\mathcal{Q}_{xz}(0,0)\right\}$ is shown. The two components of the Green's tensor vary over length scales of tens of nanometers, which is comparable to the QD size\cite{bruls02} and explains the breakdown of the dipole approximation observed in experiment.\cite{andersen10}

\begin{figure*}[t!]
	\includegraphics[width=0.8\textwidth]{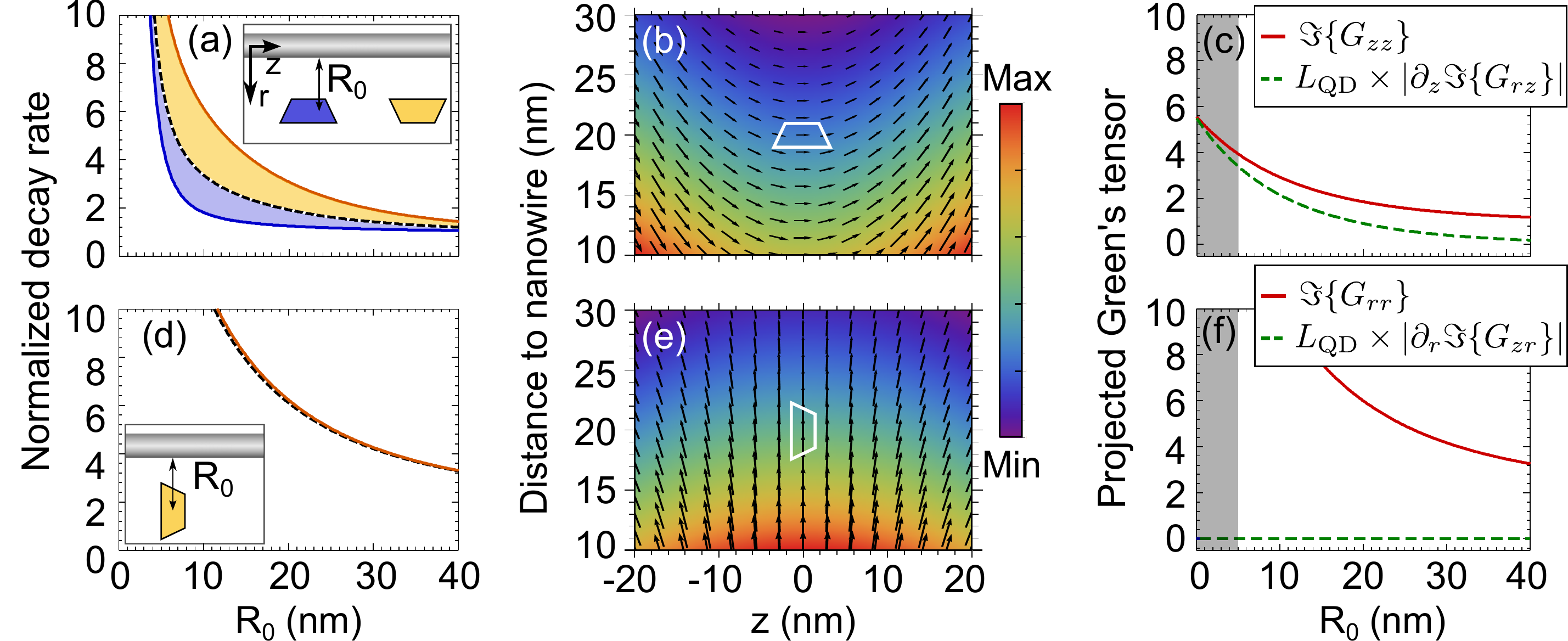}
	\caption{ \label{fig:fig3} (Color online) Probing plasmonic field gradients with mesoscopic QDs. (a) For an axially-oriented dipole, $\Gamma^{(1)}$ enhances (suppresses) the light-matter interaction strength for the configuration marked by orange (blue). The dashed line is the prediction of the dipole theory. (b) Vector plot of the plasmonic field generated by the ED of the QD situated \SI{20}{\nano\meter} away from the nanowire. Both the length of the arrows and the color scale denote the field magnitude. (c) The corresponding field projections probed by the QD and can be extracted by subtracting the decay curves in (a). The gray-shaded area is the region where nonradiative losses are dominant. (d-f) Same analysis for a radially-oriented dipole. }
\end{figure*}

QDs interact with light as spatially-extended objects and are therefore capable of probing not only the electric-field magnitude at their position (through the local density of optical states) but also field variations. This  is the basic property allowing to use QDs for probing the electromagnetic vacuum fluctuations. If placed in an unknown nanophotonic structure, the spontaneous-emission rate of the QD is generally given by $\Gamma_\blacktriangle\approx \Gamma_\blacktriangle^{(0)} + \Gamma_\blacktriangle^{(1)}$. By flipping the QD orientation, the ED contribution is the same but the first-order term changes sign, i.e., $\Gamma_\blacktriangledown\approx \Gamma_\blacktriangledown^{(0)} + \Gamma_\blacktriangledown^{(1)} = \Gamma_\blacktriangle^{(0)} - \Gamma_\blacktriangle^{(1)}$. As a consequence, both the projected Green's tensor $\Im\{ G_{xx}(0,0) \}$ and the spatial gradient $\partial_x\Im\{ G_{zx}(0,0) \}$ can be unambiguously extracted, cf. \eqref{eq:Gamma_orders}. While the former corresponds to the electric-field strength generated by an ED at the position of the emitter, the latter describes the electric-field gradient generated by the same ED. We exemplify this aspect by investigating the interaction between QDs and surface plasmons in the proximity of a silver nanowire (radius $\rho=\SI{30}{\nano\meter}$), which is capable of collecting most of the QD emission into a single propagating field mode, an important goal in the field of quantum photonics.\cite{chang07,lodahl13} From the characteristic mode equation\cite{chang07,novotny12} we find the nanowire to support a single plasmon mode with $\mathcal{G}^{(1)}=k_\mathrm{SP}\times 2\abs{\Lambda/\mu} = 0.76$. The mode is strongly confined to the nanowire surface and has a larger energy density per photon than a metal interface, which is why field gradients are further enhanced leading to a larger effect of the mesoscopic moment. The contribution of $\Gamma^{(2)}$ is again negligible since $\mathcal{G}^{(2)}=0.14$.  The coupling to radiation and ohmic-lossy modes is modelled as a point-dipole in the simple quasi-static approximation\cite{klimov04,chang07}, which gives excellent agreement with the full electrodynamic computation.\cite{chen10} In the following, only the plasmon field is discussed. The Green's tensor acquires a particularly simple form\cite{chen10} and for the geometry presented in Fig.~\ref{fig:fig3}(a) the relevant rates read
\begin{align}
\label{eq:Gamma0_Nanowire}
\frac{\Gamma_\mathrm{PL}^{(0)}}{\Gamma_\mathrm{GaAs}^{(0)}} &= C\abs{e_z(0)}^2\\
\frac{\Gamma_\mathrm{PL}^{(1)}}{\Gamma_\mathrm{GaAs}^{(0)}} &= 2C\frac{\Lambda}{\mu} \Re\left\{ \left[\partial_z e_{r}^*(0)\right]e_{z}(0)\right\},
\label{eq:Gamma1_Nanowire}
\end{align}
where $C = 3\pi c_0\epsilon_0/nk_0^2v_g$, $v_g$ is the group velocity of the guided mode and the decay rates have been normalized to the decay rate in homogeneous GaAs. $\Gamma_\mathrm{PL}^{(0)}$ is proportional to the projected energy density of the vacuum electric field and contains no phase information. On the other hand, $\Gamma_\mathrm{PL}^{(1)}$ probes the phase difference between the projected electric field and its gradient. If the electronic contribution ($\Lambda/\mu$) is in phase ($\pi$ out of phase) with the photonic contribution, the light-matter interaction is enhanced (suppressed). This is why $\Gamma^{(1)}$ vanishes for the standard-textbook plane-wave absorption,\cite{bernadotte12,tannoudji05} since a plane wave and its derivative are $\pi/2$ out of phase. Equations (\ref{eq:Gamma0_Nanowire}--\ref{eq:Gamma1_Nanowire}) contain the two field components, which can be probed by QDs in spontaneous-emission experiments, as shown in the following.

In order to acquire an understanding of the sensing capability of QDs, we analyze the properties of the surface-plasmon field. The electric field of the nanowire mode is $\mathbf{e}=\mathcal{N}\begin{pmatrix}
-\mathcal{E}_r(r), & 0, & i\mathcal{E}_z(r)
\end{pmatrix}e^{ik_\mathrm{SP}z}$,\cite{chang07} where $\mathcal{E}_r$ and $\mathcal{E}_z$ are real positive quantities and $\mathcal{N}$ a normalization constant. Thus there are two configurations in which the plasmon density of optical states is non-zero, namely for an axially- and radially-oriented dipole, see the inset of Fig.~\ref{fig:fig3}(a,c), and in the following we study the fields probed by QDs in these configurations. If the dipole moment is oriented axially, cf. Fig.~\ref{fig:fig3}(a), the first-order contribution acquires the simple form $\Gamma^{(1)}\propto -(\Lambda/\mu)k_\mathrm{SP}\mathcal{E}_r\mathcal{E}_z$ and is essentially equal in magnitude to $\Gamma^{(0)}$. This means that the coupling to surface plasmons is suppressed completely when $\Lambda$ and $\mu$ are in phase ($\Lambda/\mu>0$, depicted with blue in Fig.~\ref{fig:fig3}(a)) and enhanced by a factor of two when they are $\pi$ out of phase ($\Lambda/\mu<0$, depicted with orange). Using the aforementioned procedure of recording $\Gamma_\blacktriangle$ and $\Gamma_\blacktriangledown$ from Fig.~\ref{fig:fig3}(a), QDs can be used to probe the magnitude and curvature of the complex plasmonic field plotted in Fig.~\ref{fig:fig3}(b). At the center of the QD, the field is completely polarized along the $z$-direction and the point-dipole character of the QD therefore probes the local density of states via $\Im\{G_{zz}(0,0)\}$. Additionally, the field exhibits a curvature meaning that the radially-polarized field varies over the QD despite the fact that its mean value is zero. This radially-polarized axial gradient, $\partial_z\Im\{G_{rz}(\rr,0)\}$, is probed by the extended mesoscopic character of the QD. Both fields exhibit a monotonic increase as the QD approaches the nanowire and are plotted in Fig.~\ref{fig:fig3}(c). The axial gradient is multiplied by the in-plane QD size $(L_\mathrm{QD}=\SI{20}{\nano\meter})$\cite{bruls02} to show the field variation over the QD spatial extent. It is interesting to note that the field $\Im\{G_{rz}\}$ exhibits a large variation over the QD that is comparable to the probed field itself $\Im\{G_{zz}\}$. This example clearly shows the "ease" of breaking the dipole approximation with mesoscopic QDs in nanophotonic structures. We find that essentially the entire contribution to $\Gamma_\mathrm{PL}^{(1)}$ stems from the EQ nature of the QD, in contrast to the silver interface, where the MD and EQ contributions are of comparable magnitude. These examples show that, even though the MD and EQ moments of the QD are equal in magnitude, their individual contribution to the light-matter interaction can be tailored by correspondingly engineering the nanophotonic environment. In this sense, QDs are promising light emitters for embedment in optical metamaterials, whose practical realization has become technologically feasible over the past years.

In the second configuration, the dipole moment is oriented radially and the first-order contribution $\Gamma^{(1)}$ vanishes because the environment is parity-symmetric along the QD height, see the inset of Fig.~\ref{fig:fig3}(d). Consequently, the dipole approximation is a very good assumption for this configuration. As seen in Fig.~\ref{fig:fig3}(d), the prediction of the two theories are very close. For a better understanding, we plot the in-phase component of the electric field generated by the dipole character of the QD in Fig.~\ref{fig:fig3}(e). The mesoscopic moment $\Lambda\equiv \Lambda_{rz}$ would couple to the radial gradient of the $z$-polarized field but the latter vanishes in this configuration owing to the aforementioned parity symmetry. There are two field gradients that do not vanish, the $z$-derivative of the $z$-polarized field and the $r$-derivative of the $r$-polarized field.  However, they are not sensed by QDs because they couple to other mesoscopic moments ($\Lambda_{zz}$ and $\Lambda_{rr}$, respectively), which vanish for In(Ga)As QDs. Therefore, a radial QD probes only the (electric) local density of optical states as illustrated in Fig.~\ref{fig:fig3}(f).

In conclusion, we have shown that the commonly employed In(Ga)As QDs are capable of strongly interacting with the multipolar quantum vacuum on dipole-allowed transitions. This striking behavior is triggered by the lack of parity-symmetry of the electronic wavefunctions in the growth direction, a feature that is absent in atomic physics because atoms have parity symmetry. The first-order expansion term $\Gamma^{(1)}$ can be comparable in magnitude to the dipole rate $\Gamma^{(0)}$ in nanophotonic structures. This effect can be exploited to use QDs as a probe of the local field environment revealing not only information about the field itself but also about its gradients. Furthermore, by engineering the nanophotonic environment it is possible to selectively access the MD or EQ nature of the QD and, thereby, to tailor the multipolar radiation of semiconductor QDs. We have exemplified this for metal nanostructures but any strongly- or rapidly-varying optical modes would produce deviations from the dipole approximation and we therefore expect this work to be of significance not only for plasmon-based devices\cite{schuller10} and photovoltaics,\cite{atwater10} but also for the active field of photonic-crystal cavities and waveguides, where QDs have been described as dipole emitters so far.

\section*{Acknowledgements}
We thank P.~T.~Kristensen for valuable discussions. We gratefully acknowledge the financial support from the Danish Council for Independent Research (natural sciences and technology and production sciences), the European Research Council (ERC consolidator grants "ALLQUANTUM" and "QIOS"), and the Carlsberg Foundation.

\bibliographystyle{nature}
\bibliography{bibliography}

\begin{thebibliography}{10}

\bibitem{noecker88}
Noecker, M., Masterson, B., and Wieman, C.
\newblock {\em Phys. Rev. Lett.}{ \bf 61}, 310 (1988).

\bibitem{rukhlenko09}
Rukhlenko, I.~D., Handapangoda, D., Premaratne, M., Fedorov, A.~V., Baranov,
  A.~V., and Jagadish, C.
\newblock {\em Opt. Express}{ \bf 17}, 17570 (2009).

\bibitem{taminiau12}
Taminiau, T.~H., Karaveli, S., van Hulst, N.~F., and Zia, R.
\newblock {\em Nat. Commun.}{ \bf 3}, 979 (2012).

\bibitem{zurita01}
Zurita-S{\'a}nchez, J.~R. and Novotny, L.
\newblock {\em JOSA B}{ \bf 19}, 1355 (2002).

\bibitem{zurita02}
Zurita-S{\'a}nchez, J.~R. and Novotny, L.
\newblock {\em JOSA B}{ \bf 19}, 2722 (2002).

\bibitem{andersen10}
Andersen, M.~L., Stobbe, S., S{\o}rensen, A.~S., and Lodahl, P.
\newblock {\em Nat. Phys.}{ \bf 7}, 215 (2011).

\bibitem{wood97}
Wood, C., Bennett, S., Cho, D., Masterson, B., Roberts, J., Tanner, C., and
  Wieman, C.
\newblock {\em Science}{ \bf 275}, 1759 (1997).

\bibitem{ciraci12}
Cirac{\`\i}, C., Hill, R., Mock, J., Urzhumov, Y.,
  Fern{\'a}ndez-Dom{\'\i}nguez, A., Maier, S., Pendry, J., Chilkoti, A., and
  Smith, D.
\newblock {\em Science}{ \bf 337}, 1072 (2012).

\bibitem{toscano13}
Toscano, G., Raza, S., Yan, W., Jeppesen, C., Xiao, S., Wubs, M., Jauho, A.-P.,
  Bozhevolnyi, S.~I., and Mortensen, N.~A.
\newblock {\em Nanophotonics}{ \bf 2}, 161 (2013).

\bibitem{bernadotte12}
Bernadotte, S., Atkins, A.~J., and Jacob, C.~R.
\newblock {\em J. Chem. Phys.}{ \bf 137}, 204106 (2012).

\bibitem{soukoulis10}
Soukoulis, C.~M. and Wegener, M.
\newblock {\em Science}{ \bf 330}, 1633 (2010).

\bibitem{pendry06}
Pendry, J.~B., Schurig, D., and Smith, D.~R.
\newblock {\em Science}{ \bf 312}, 1780 (2006).

\bibitem{burresi09}
Burresi, M., Van~Oosten, D., Kampfrath, T., Schoenmaker, H., Heideman, R.,
  Leinse, A., and Kuipers, L.
\newblock {\em Science}{ \bf 326}, 550 (2009).

\bibitem{lodahl13}
Lodahl, P., Mahmoodian, S., and Stobbe, S.
\newblock {\em arXiv/1312.1079}{ \bf } (2013).

\bibitem{wang11}
Wang, Q., Stobbe, S., and Lodahl, P.
\newblock {\em Phys. Rev. Lett.}{ \bf 107}, 167404 (2011).

\bibitem{tighineanu13}
Tighineanu, P., Daveau, R., Lee, E.~H., Song, J.~D., Stobbe, S., and Lodahl, P.
\newblock {\em Phys. Rev. B}{ \bf 88}, 155320 (2013).

\bibitem{tannoudji05}
Cohen-Tannoudji, C., Diu, B., and Lalo{\"e}, F.
\newblock {\em {Quantum Mechanics}}, volume~2.
\newblock John Wiley \& Sons,  (2005).

\bibitem{vats02}
Vats, N., John, S., and Busch, K.
\newblock {\em Phys. Rev. A}{ \bf 65}, 043808 (2002).

\bibitem{novotny12}
Novotny, L. and Hecht, B.
\newblock {\em Principles of nano-optics}.
\newblock Cambridge University Press,  (2012).

\bibitem{ahn03}
Jun~Ahn, K. and Knorr, A.
\newblock {\em Phys. Rev. B}{ \bf 68}, 161307 (2003).

\bibitem{stobbe12}
Stobbe, S., Kristensen, P.~T., Mortensen, J.~E., Hvam, J.~M., M\o{}rk, J., and
  Lodahl, P.
\newblock {\em Phys. Rev. B}{ \bf 86}, 085304 (2012).

\bibitem{kristensen13}
Kristensen, P.~T., Mortensen, J.~E., Lodahl, P., and Stobbe, S.
\newblock {\em Phys. Rev. B}{ \bf 88}, 205308 (2013).

\bibitem{paulus00}
Paulus, M., Gay-Balmaz, P., and Martin, O.~J.
\newblock {\em Phys. Rev. E}{ \bf 62}, 5797 (2000).

\bibitem{drexhage70}
Drexhage, K.
\newblock {\em J. Lum.}{ \bf 1}, 693 (1970).

\bibitem{tighineanu14}
Tighineanu, P., S{\o}rensen, A.~S., Stobbe, S., and Lodahl, P.
\newblock {\em to be submitted}{ \bf } (2014).

\bibitem{bruls02}
Bruls, D., Vugs, J., Koenraad, P., Salemink, H., Wolter, J., Hopkinson, M.,
  Skolnick, M., Long, F., and Gill, S.
\newblock {\em Appl. Phys. Lett.}{ \bf 81}, 1708 (2002).

\bibitem{chang07}
Chang, D.~E., S\o{}rensen, A.~S., Hemmer, P.~R., and Lukin, M.~D.
\newblock {\em Phys. Rev. B}{ \bf 76}, 035420 (2007).

\bibitem{klimov04}
Klimov, V.~V. and Ducloy, M.
\newblock {\em Phys. Rev. A}{ \bf 69}, 013812 (2004).

\bibitem{chen10}
Chen, Y., Nielsen, T.~R., Gregersen, N., Lodahl, P., and M{\o}rk, J.
\newblock {\em Phys. Rev. B}{ \bf 81}, 125431 (2010).

\bibitem{schuller10}
Schuller, J.~A., Barnard, E.~S., Cai, W., Jun, Y.~C., White, J.~S., and
  Brongersma, M.~L.
\newblock {\em Nat. Mat.}{ \bf 9}, 193 (2010).

\bibitem{atwater10}
Atwater, H.~A. and Polman, A.
\newblock {\em Nat. Mat.}{ \bf 9}, 205 (2010).

\bibitem{moison94}
Moison, J., Houzay, F., Barthe, F., Leprince, L., Andre, E., and Vatel, O.
\newblock {\em Appl. Phys. Lett.}{ \bf 64}, 196 (1994).

\bibitem{debeer08}
DeBeer~George, S., Petrenko, T., and Neese, F.
\newblock {\em Inorganica Chimica Acta}{ \bf 361}, 965 (2008).

\end{thebibliography}

\section*{Supplementary Information}
\subsection*{1. Symmetry analysis of the mesoscopic moments}

The number of non-zero entries in $\mathbf{\Lambda}$ and $\mathbf{\Omega}$ can be greatly reduced by employing simple parity considerations. We assume the QDs to be lens-shaped with in-plane cylindrical symmetry but with no parity symmetry in the growth direction, in good agreement with the shape of In(Ga)As QDs.\cite{moison94} Importantly, this analysis is not necessarily bound to this particular wavefunction shape and is also valid for pyramidal or in-plane elliptical QDs. In the effective-mass approximation, the single-particle wavefunctions are given by $\Psi(\rr)=u(\rr)\psi(\rr)$, where $u$ is the periodic Bloch function at the center of the Brillouin zone and $\psi$ the slowly-varying envelope. Due to the exchange interaction, the two bright excitons are linearly polarized along $x$ and $y$, respectively. Therefore, we consider only one of the bright excitons $u_g=u_x$ but the results are analogous for the other one.

The valence-band Bloch function $u_x$ inherits the symmetry of the $p_x$ orbital and therefore exhibits odd parity ("-1") in the $x$-direction and even parity ("+1") in $y$ and $z$. The conduction-band Bloch function $u_e$ inherits the spherical symmetry of the $s$-orbital and therefore contains even parity in all directions.  Since the effective-mass theory is an envelope-function formalism, the slowly-varying envelopes $\psi$ inherit the QD symmetry. Table~\ref{table:parityWF} summarizes these considerations. Applying parity-symmetry arguments, we find that only $\Lambda_{xz}=\bracket{\Psi_g}{x\hat{p}_z}{\Psi_e}$ and $\Lambda_{zx}=\bracket{\Psi_g}{z\hat{p}_x}{\Psi_e}$ contain non-zero entries. Below we argue that the contribution of $\Lambda_{zx}$ is negligible for In(Ga)As QDs, as is the contribution of the entries in $\mathbf{\Omega}$. In general, the value of the mesoscopic moments $\Lambda=\Lambda_{xz}$ and $\Lambda_{zx}$ depends on the choice of the coordinate system. Here, the in-plane cylindrical symmetry of the QD provides a natural choice of the origin of the coordinate system, which renders $\Lambda$ a well-defined QD parameter. The QD therefore exhibits two degrees of freedom while interacting with light, one originating from the dipole moment $\mu$ and the other from the mesoscopic moment $\Lambda$.
\begin{table}[H]
\begin{center}
\begin{tabular}{c | c c c | c c c}\botrule
                   & $u_g$            & $\psi_{g}$     & $\Psi_{g}$    & $u_e$         & $\psi_e$  & $\Psi_e$  \\
$x$				   & -1               & 1              & -1            & 1             & 1           & 1      \\
$y$                & 1                & 1              & 1             & 1             & 1           & 1      \\
$z$                & 1                & 0              & 0             & 1             & 0           & 0      \\
\botrule
\end{tabular}
\end{center}
\caption{Symmetries of the electron and hole wavefunctions for a lens-shaped QD. "1" denotes even parity, "-1" odd parity and "0" no parity.}
\label{table:parityWF}
\end{table}

\subsection*{2. Justification for neglecting $\boldsymbol \Lambda_{\boldsymbol z \boldsymbol x}$}
\begin{figure}[H]
\center
\includegraphics[width=0.35\textwidth]{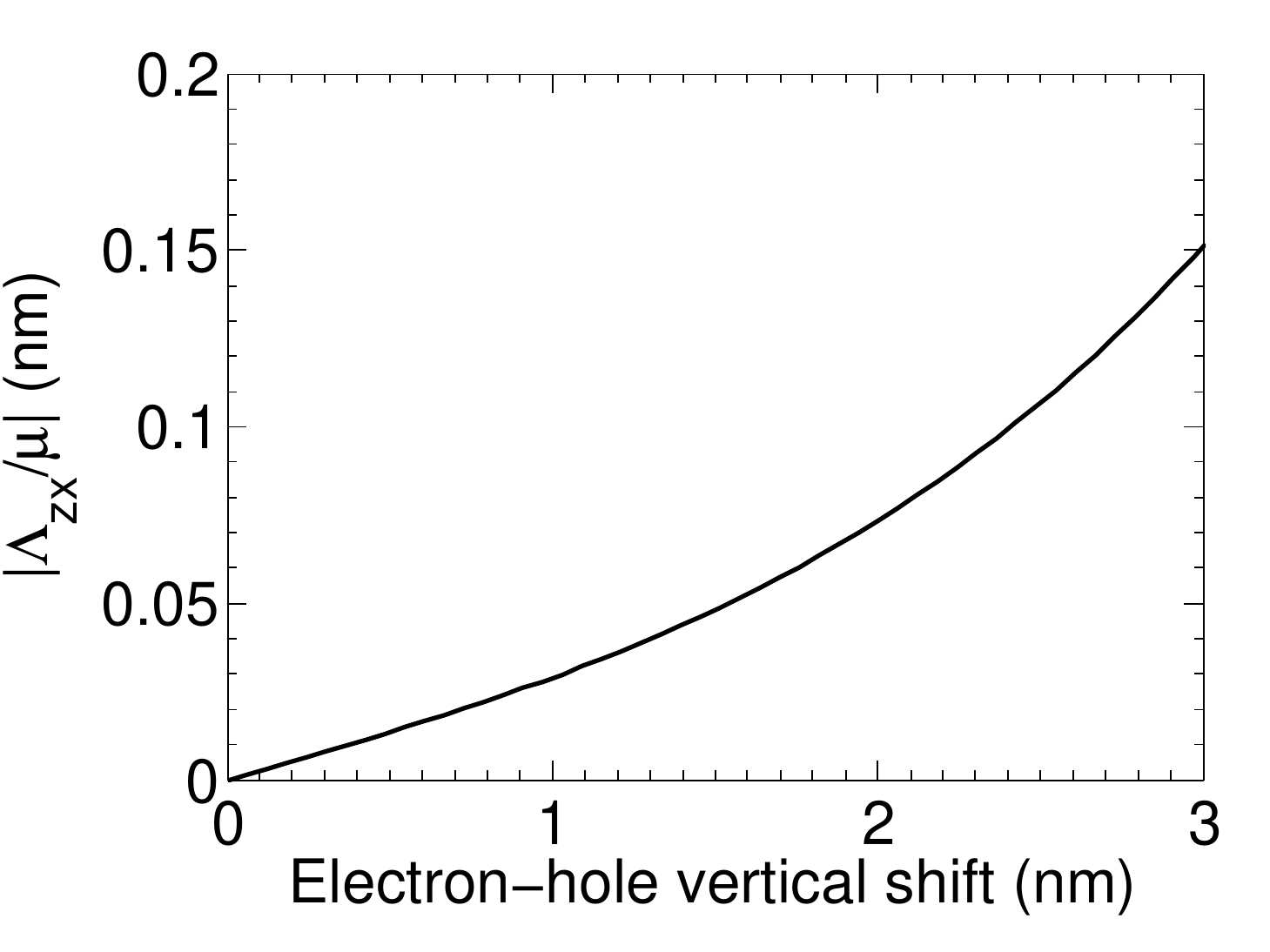}
\caption{$\abs{\Lambda_{zx}/\mu}$ as a function of the electron-hole vertical shift.}
\label{fig:Lambda_zx}
\end{figure}

Here we give an estimate for the first-order mesoscopic moment $\Lambda_{zx}$ and show that it has a negligible contribution to the light-matter interaction. Recall that
\begin{equation}
\Lambda_{zx} = \bra{\Psi_{g}} (z-z_0) \hat{p}_x \ket{\Psi_e} \approx p_{cv}\bra{\psi_{g}} (z-z_0) \ket{\psi_e},
\end{equation}
where $p_{cv}$ is the Bloch matrix element. The relevant figure of merit for the magnitude of the mesoscopic moment is
\begin{equation}
\abs{\frac{\Lambda_{zx}}{\mu}} = \abs{\frac{\bra{\psi_{g}} (z-z_0) \ket{\psi_e}}{\ovI{\psi_{g}}{\psi_e}}} = \abs{\frac{\bra{\psi_{g}} z \ket{\psi_e}}{\ovI{\psi_{g}}{\psi_e}}-z_0}
\end{equation}
The QD is not symmetric in the $z$-direction and there is no predetermined choice for $z_0$. Conceptually, $z_0$ should be defined such that the expansion in $\Gamma$ converges fastest. As shown in Ref.~\citenum{debeer08}, this is the case if $z_0$ corresponds to the region where the transition density is of the largest absolute value. For an exciton, the center-of-mass coordinate has the largest transition density and therefore
\begin{equation}
z_0 = \frac{z_e+m_rz_{g}}{1+m_r},
\end{equation}
where $m_r$ is the ratio of the effective masses of the hole and the electron. Now we are in a position to estimate the magnitude of $\Lambda_{zx}$. Since this is an effect involving the slowly-varying envelopes, we can make some realistic assumptions. We assume Gaussian wavefunctions for the electron and hole with an out-of-plane HWHM of \SI{2}{\nano\meter} for the electron and $\sqrt{\xi}$ smaller for the hole, where $\xi=5$ is the ratio of their effective masses. Then, we plot $\abs{\Lambda_{zx}/\mu}$ as a function of the distance between the electron and hole wavefunctions, see \figref{fig:Lambda_zx}.
For a realistic vertical shift of 2--\SI{3}{\nano\meter} we obtain $\abs{\Lambda_{zx}/\mu}\approx \SI{0.1}{\nano\meter}$. Then, the relevant figure of merit for the breakdown of the dipole approximation is $2k\abs{\Lambda_{zx}/\mu} \approx 0.2\% \ll 1$. A more general explanation of why $\Lambda_{zx}$ is negligible is related to the small QD height of about 4 nanometers.\cite{bruls02} This analysis provides rigorous justification for neglecting $\Lambda_{zx}$ both in this work and in Ref.~\citenum{andersen10}.

\subsection*{3. Justification for neglecting $\boldsymbol \Omega$}
The lowest-order contribution of the second-order mesoscopic moment $\Omega_{ijk}$ to the light-matter interaction is $\Gamma^{(2)}$. Therefore, its relevance with respect to $\Gamma^{(0)}$ can be defined heuristically via the magnitude of $k^2L_\mathrm{QD}^2$. In this work, $kL_\mathrm{QD} \sim 0.3$, which implies a value of $k^2L_\mathrm{QD}^2$ below 0.1. Since we investigate dipole-allowed transitions where both $\Gamma^{(0)}\neq 0$ and $\Gamma^{(1)}\neq 0$, we neglect the contribution of $\mathbf{\Omega}$ in this work.

\end{document}